\documentstyle[epsfig]{l-aa}

\begin{document}

\thesaurus{07(08.03.4;08.05.2;08.02.1;08.14.1;13.25.5)}

\title{Large-scale perturbations in the circumstellar envelopes of 
Be/X-ray binaries}

\author{I.~Negueruela\inst{1,2}
\and P.~Reig\inst{2}
\and M.~J.~Coe\inst{2}
\and J.~Fabregat\inst{3}
}

\offprints{I. Negueruela (Liverpool address)}

\institute{Astrophysics Research Institute, Liverpool John Moores University,
Byrom St., Liverpool, L3 3AF, U.K.
\and Physics and Astronomy Department, Southampton University,
Southampton, SO17 1BJ, U.K.
\and Departamento de Astronom\'{\i}a y Astrof\'{\i}sica, Universidad
de Valencia, 46100, Burjassot, Valencia, Spain}

\date{Received 7 July 1997    / Accepted    April 1998 }

\maketitle 

\markboth{Negueruela et al. : Circumstellar envelopes of Be/X-ray binaries}
{Negueruela et al. : Circumstellar envelopes of Be/X-ray binaries}

\begin{abstract}

We investigate the spectroscopic characteristics of the optical 
components of Be/X-ray binary systems, using data collected
during our seven-year monitoring campaign. We find examples of major
changes in the emission line profiles associated with Type II X-ray
outbursts, later developing into V/R variability cycles. We show that the
time-scales for V/R variability in Be/X-ray transients extend from a 
few weeks to years and interpret all these changes as due to the 
presence of global disruptions of the axisymmetric density distribution
in the extended envelopes of the Be stars in these systems. The association 
between X-ray
outbursts and V/R variability, the occurrence of very fast changes and
the very short quasi-periods of variability displayed by Be/X-ray binaries
lead us to conclude that the presence of the neutron star is an important 
factor
affecting the dynamics of the disc-like envelopes. The interaction 
between the compact companion and the disc would explain
the correlation between H$\alpha$ strength and orbital period recently
found. The characteristics of the V/R cycles are, however, mainly independent 
of 
the binary parameters.


\end{abstract}

\section{Introduction}

Be/X-ray binaries constitute the major subclass of massive X-ray
binaries, in which  X-ray emission is due to accretion
of matter from an early-type  mass-losing star by a compact
companion (see Apparao 1994, White et al. 1995, for reviews). Be stars are 
early-type  non-supergiant stars, which at some
time have shown emission in the Balmer lines. Both the emission
lines and the characteristic strong infrared excess when compared to
normal stars of the same spectral types are attributed to the presence
of a cool circumstellar envelope, presumably in the shape of a disc
(see Slettebak 1988). The physical reasons which give rise to the disc
are unknown, but it is generally believed that the high rotational velocity 
of Be stars plays an important role, even though it is accepted that some
other mechanism(s) must be at work. 

Most Be/X-ray binaries have relatively eccentric orbits and the compact
companion (in general, a neutron star, but in some cases possibly a white
dwarf) spends most of its time far away from the disc surrounding the Be 
star. Three kinds of X-ray activity are observed (Stella et al. 1986, 
henceforth SWR):

\begin{enumerate}
\renewcommand{\theenumi}{(\arabic{enumi})}
\item Persistent low-luminosity ($L_{{\mathrm x}} \la 10^{36}$ erg s$^{-1}$) 
X-ray emission. Some sources (e.g., X Persei) have always been 
observed in this state.

\item Periodical (Type I in SWR) X-ray outbursts ($L_{{\mathrm x}} 
\approx 10^{36} - 10^{37}$ erg s$^{-1}$), coinciding with 
the periastron passage of the neutron star. 
Type I outbursts have been observed in numerous sources, such as
A\,0535+262 (Motch et al. 1991) and EXO\,2030+375 (Norton et 
al. 1994).

\item Giant (Type II in SWR) X-ray outbursts ($L_{{\mathrm x}} \ga 10^{37}$ erg 
s$^{-1}$), which do not show any orbital modulation. Type II outbursts are 
normally seen in those sources that also display Type I activity (see 
 Parmar et al. 1989, Finger et al. 1996a for examples).
\end{enumerate}

Be/X-ray binary systems which display outbursts are collectively termed 
Be/X-ray transients. Most transients (e.g. A\,0535+26) also show 
low-luminosity X-ray emission when they are not in outburst, but in systems
containing fast-rotating neutron stars, centrifugal inhibition of 
accretion prevents X-ray emission (SWR) except during outbursts (e.g.,
4U\,0115+634). Type I outbursts occur in series between long periods of 
X-ray inactivity (or low-luminosity emission), while the onset of
Type II outbursts is completely unpredictable.

\section{The circumstellar envelopes of Be/X-ray binaries}

\subsection{Line profiles and circumstellar structure}

Loose correlations between the optical/infrared properties of Be/X-ray
binaries and their X-ray behaviour have been observed to exist (e. g.,
Corbet et al. 1986a; Coe et al. 1994). This correlation is to be expected, 
since 
the optical/infrared observations provide information about the changing 
conditions in the circumstellar envelope from which the neutron star is 
accreting. Simple models of the circumstellar environment have
been used to fit the observed lightcurves (see Apparao 1994 and references
therein). In the most basic of these models, the size of the disc is
supposed to be the main factor. When the disc is so large that it reaches the
orbit of the neutron star, Type I X-ray outbursts occur. When the disc
is smaller the neutron star cannot accrete. Low-luminosity emission can
be due to accretion from the low-density transition regions between the 
envelope and the interstellar material or the low-density fast wind 
believed to be emitted from the polar regions of Be stars (Lamers \& 
Waters 1987, Slettebak 1988).

Waters et al. (\cite{waters88}) showed that the infrared photometric 
magnitudes and the X-ray lightcurves of Be/X-ray binaries indicated 
that the neutron stars 
were accreting from a high-density, low-velocity wind . Waters et al. 
(\cite{waters89}) 
analysed the influence of the changing conditions in the circumstellar 
envelope on the X-ray lightcurves making use of a more complicated model, 
which takes into account the rotation of 
the envelope. They found that wind velocities in the 
range 100 --- 600 km s$^{-1}$ at the distance of periastron passage of the
neutron star can account for the observed X-ray luminosities. In their
model, the relative velocity of the wind was the main factor affecting the
X-ray luminosity. High 
luminosities during Type II outbursts imply small relative velocities 
($\sim 100$ km s$^{-1}$), while Type I outbursts imply larger velocities.

However, the existence of this wind outflow at large distances --- which 
would be common to all Be 
stars --- is not reflected in the shapes of the H$\alpha$ emission lines, 
which are frequently symmetric and believed to be determined by rotation 
(Hanuschik et al. 1993). Therefore it seems that a heretofore unknown 
mechanism (see Chen et al. 1992 for a discussion) accelerates the 
circumstellar material 
outwards in the regions beyond the H$\alpha$ formation zone. 
The exact size and location of this zone is not known, but most estimates 
support an outer radius in the range 5 --- 20 $R_{*}$ (see Hummel \& 
Vrancken  1995).

In recent years, a much improved description of the structure of the
circumstellar envelopes of Be stars has begun to emerge. The appearance 
of high-resolution spectral atlases (Dachs et al. 1992, Hanuschik et al. 
1996) has resulted in great advances in the traditional analysis of the
emission line profiles (Slettebak et al. 1992, Hummel \& Vrancken 1995). 
The main conclusion reached is that, in spite of the multiplicity of line
shapes (single, double or triple-peaked, with or without flank inflections,
showing  central self-absorption reversal and/or emission wings, etc) the
most fundamental division of emission profiles in Be stars can be made into 
two main categories (Hanuschik et al. 1995): 

\begin{enumerate}
\renewcommand{\theenumi}{(\roman{enumi})}

\item Symmetric profiles, generally presenting two peaks of similar intensity. 
These shapes only evolve very slowly (with time-scales of a few years) and can 
remain unchanged for years.

\item Asymmetric profiles with a higher degree of variability. These 
profiles undergo quasi-cyclic V/R variability (the ratio between the V(iolet)
and the R(ed) peak changes regularly), with quasi-periods ranging 
from a few years to decades. 

\end{enumerate}

The symmetric profiles are believed to arise from stable quasi-Keplerian 
discs, while the asymmetric shapes are produced in discs with a perturbed 
density distribution.
The  perturbations are associated with the existence of global oscillation 
modes propagating through the envelopes (Okazaki 1991, 1996; Papaloizou et 
al. 1992). The only modes which can propagate in a nearly Keplerian disc are 
global $m = 1$ (where $m$ is the azimuthal wave number)
oscillations (Kato 1983). Theoretical line 
profiles calculated from models of quasi-Keplerian discs with $m =1$ global
oscillation modes are in agreement with observed profiles (Hummel \& Hanuschik
1997, Hummel \& Vrancken 1995, Okazaki 1996). The evolution of observed V/R 
variations in some Be stars can be readily explained by a progressing global 
mode (Telting et al. 1994, Reig et al. 1997a). As a consequence, it is now 
generally accepted that asymmetric line profiles in the spectra of Be stars
are caused by an asymmetric matter configuration in their extended envelopes,
due to the existence of progressing density waves (Hummel \& Hanuschik 1997, 
Okazaki 1997). Hanuschik (1996) indicates that approximately two thirds of the 
bright Be stars in his sample show symmetric profiles at a given 
time. However, evolution is observed over long time-scales, and approximately
two thirds of the Be stars which have been extensively monitored have displayed
V/R variability at some time (Okazaki 1997).

\subsection{Observations of Be/X-ray binaries}

It has been traditionally believed that the presence of the neutron star
in Be/X-ray binaries will not affect the dynamics of the Be envelope. 
Norton et al. (\cite{norton}) could not detect any variability in the 
photometric 
or spectroscopic properties of the Be/X-ray
system EXO\,2030+375 during a Type I outburst. They deduced that the
effects of the compact companion in the structure of the circumstellar
envelope were, in general, negligible.
Recently, however, it has been proposed (Reig et al.
1997b) that the optical components of Be/X-ray binaries form a distinct
group inside the Be stars. Supporting this hypothesis, Reig et
al. (\cite{reigb}) call two observational facts. First, there seems to be a
correlation between the maximum equivalent width (EW) of the H$\alpha$
line which a system has shown and the orbital period of the neutron
star. Second, Be stars forming part of Be/X-ray binaries have, on
average, low H$\alpha$ EW when compared to randomly selected sets of
Be stars.

The presence of strongly asymmetric lines in the spectra of Be/X-ray binaries, 
similar to those observed in Be stars undergoing V/R variability has been 
noted in several occasions (e.g., Cook \& Warwick 1987, Corbet et al. 1986a). 
However, 
the lack of long-term monitoring of the sources has not permitted to determine 
whether these asymmetric profiles were associated with cyclic changes or were 
due to some other process. In these complex systems, the presence of temporary 
sources of H$\alpha$ emission, such as an accretion disc or a high-ionization 
region, heated up by X-rays, around the neutron star cannot 
be ruled out. These additional sources of H$\alpha$ emission would add to the 
circumstellar emission and create complicated line profiles.

Until now, the evolution of emission lines during Type II events
had not been studied due to their unpredictability. No observations were taken 
at the time of the only 
Type II outburst ever shown by EXO\,2030+375 (Parmar et al. 1989) or 
2S\,1417$-$624 (Finger et al. 1996b) or any of the two Type II 
outbursts of V\,0332+53 (Terrell \& Priedhorsky 
1984; Takeshima et al. 1994). Likewise, no optical coverage 
of the 1973 Type II outbursts of 4U\,1145$-$619 exists. A bright
outburst from this source in 1994 could have been of Type II. A strong 
disturbance of the envelope is discussed by Stevens et al. (1997), but their
sparse observations do not allow us
to determine if its V/R periodicity was affected.
Asymmetric lines have been observed in the spectra of A\,0535+26 on 
different occasions, but its previous Type II outbursts in 1975 and 1980 were 
not covered. The spectra of V635 Cas (4U\,0115+634) during and after another 
Type II 
outburst in 1980 probably show asymmetric H$\alpha$ profiles (Kriss et al. 
1983), but their very low S/N ratio does not allow us to be certain. No 
evident changes in symmetry were associated with a Type II X-ray outburst in 
1991 (Negueruela et al. 1997). 

In this paper we show that Be/X-ray binaries display quasi-cyclic changes 
similar to those observed in isolated Be stars and present evidence indicating 
that Type II X-ray outbursts affect the V/R variability of the optical 
components. Two sources which have recently displayed X-ray activity are 
investigated in detail:

\subsubsection{4U\,0115+634}

The hard X-ray transient 4U\,0115+634 is one of the best studied 
Be/X-ray binary systems (see Campana 1996; Negueruela et al. 1997, henceforth 
N97). It consists of a fast-rotating ($P_{{\mathrm s}} = 3.6 \: {\mathrm s}$) 
neutron star in a relatively close ($P_{{\mathrm orb}} = 24.3 \: {\mathrm d}$) 
and eccentric ($e = 0.34$) orbit around the O9.5Ve star V635 Cas (see 
Tamura et al. 1992; Unger et al. 1997). Due to the fast rotation of the 
neutron star, centrifugal inhibition of 
accretion prevents the onset of Type I outbursts (SWR, N97). The system is 
the only known Be/X-ray transient that has solely been observed to display 
Type II activity, with long strong X-ray outbursts extending over more than 
one orbital period and showing no dependence on any orbital parameters. 
These giant 
outbursts are believed to be due to episodes of enhanced mass loss from the Be 
star, which are reflected in optical and infrared brightening events (SWR, 
N97).

A Type II X-ray outburst from 4U\,0115+634 was detected by the BATSE 
experiment on board the {\em CGRO} satellite 
starting on Nov. 18, 1995  (Finger et al. 1995). Granat/{\em WATCH} 
measured the flux in the 8--20 keV band peak at $670 \pm 60$ mCrab on Nov. 
25 (Sazonov \& Sunyaev 1995), which remained at that level until early
December. A second flare lasted into January 1996 (Scott et al. 1996).
The outburst was the strongest during the last decade (see the X-ray 
lightcurve in N97).

\begin{figure*}[ht]
\begin{center}
   \leavevmode
\epsfig{file=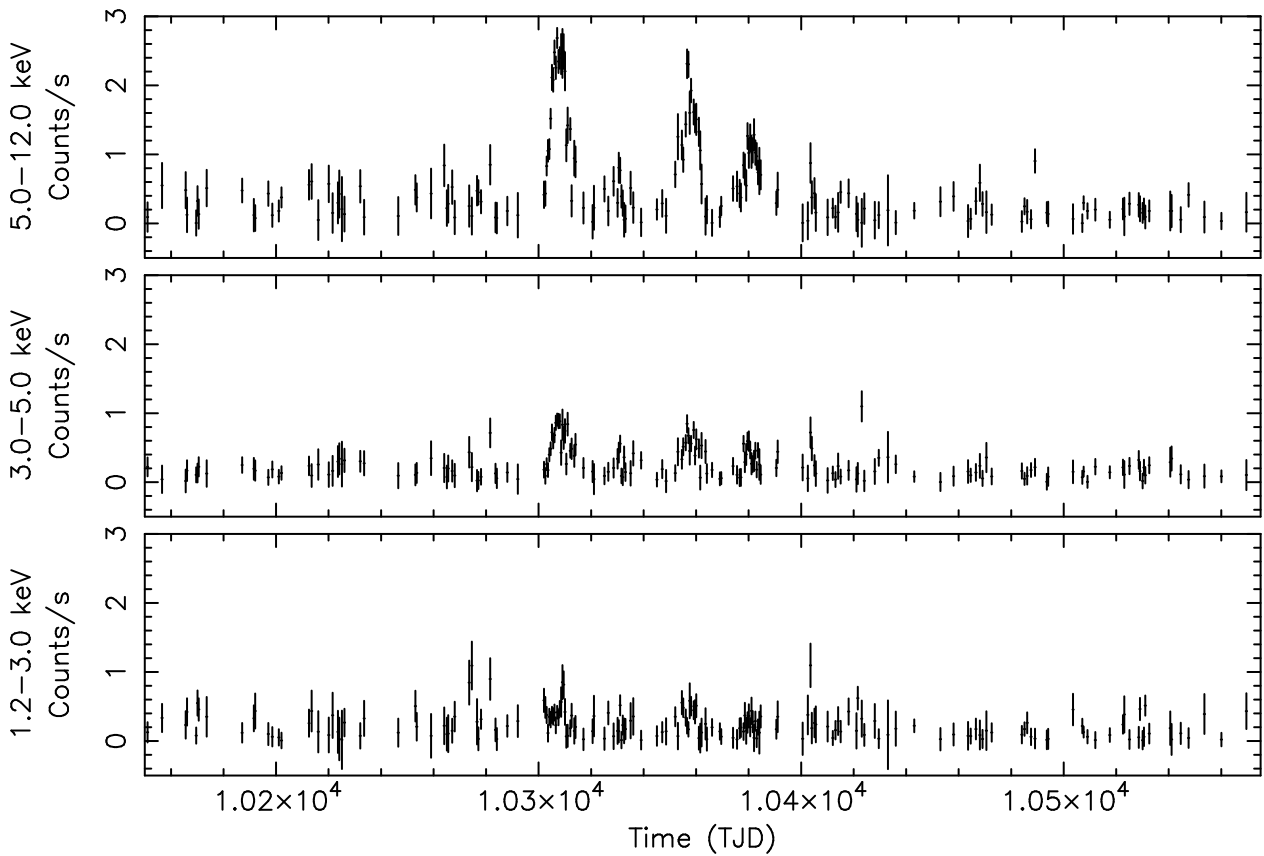, width=16.8cm, bbllx=10pt,
 bblly=270pt, bburx=496pt, bbury=600pt, clip=}
\end{center}
\caption{X-ray lightcurves of the Be/X-ray binary 4U\,0115+634 in three 
energy bands ($1.3\,-\,3.0$, $3.0\,- \,5.0$ and $5.0\,-\, 12.2$ keV), taken 
with the 
All Sky Monitor on board {\em RXTE}. Points represent 18-h averages of the 
individual dwell solutions.}
\label{fig:xray}
\end{figure*}

\subsubsection{A\,0535+262}

The Be/X-ray transient A\,0535+262 consists of a neutron star orbiting 
the O9.7IIIe star HD 245770 in a wide and eccentric orbit ($P_{{\mathrm 
orb}} = 110.3$ d, $e = 0.47$, Finger et al. 1994, 1996a). Due to its 
brightness, 
HD 245770 has been extensively monitored (see Motch et al. 1991, 
Giovannelli \& Graziati 1992 for reviews). Clark et al. (1998a, henceforth 
Cl98) 
have presented the results of several years 
of multiwavelength observations and analysed the connection between the X-ray 
activity and the behaviour of the primary. They did not find any clear 
correlation. The source displays all three types of X-ray activities observed 
in Be/X-ray binaries. The X-ray flux in the 2 -- 10 keV is normally in the 
range 5 -- 10 mCrab during quiescence phases. Type I outbursts occur close 
to the periastron passage of the neutron star, but Type II outbursts 
($F_{{\mathrm x}} \ga 1$ Crab) are generally slightly delayed in phase. 
The last active period of 
the source occurred between March 1993 and 
September 1994. Type I outbursts were observed at all periastron passages, 
except on February 1994, when a Type II outburst took place (Finger et al. 
1996a). The Type II outburst lasted 52 days and peaked on February 18, when it 
reached a 
flux of 8 Crab in the 20 -- 40 keV band.

\section{Observations}
\label{sec:observ}
 
As part of the Southampton/Valencia/SAAO long-term monitoring campaign of 
Be/X-ray binaries, we have obtained H$\alpha$ spectroscopy of a number of 
sources. The details of the programme are described in Reig et al. (1997c). 
Here 
we concentrate on the temporal evolution of the H$\alpha$ line profile in 
search 
of the existence of V/R variability in these systems and the characteristics 
of this variability. Most of the data presented here has not been published 
previously, though some spectra have appeared in a previous paper (Clark et 
al. 1998a). The spectra have been obtained with the 2.5-m 
Isaac Newton Telescope (INT), the 4.2-m William Herschel Telescope (WHT) and 
the 1.0-m Jakobus Kapteyn Telescope (JKT), all three located at the 
Observatorio 
del Roque de los Muchachos, La Palma, Spain, and the 1.5-m telescope at 
Palomar Mountain (PAL). Different telescope configurations were used on 
different dates, 
some of which are discussed in the following sections.

\subsection{Observations of 4U\,0115+634 (V635 Cas)}

\subsubsection{X-ray observations}

The All Sky Monitor (ASM) on board the Rossi X-ray Timing Explorer {\em RXTE} 
satellite consists of three wide-angle Scanning Shadow Cameras (SSCs) mounted 
on 
a rotating drive assembly, which scan $\sim 70$ \% of the sky every 1.5 hours. 
A 
description of the satellite and its data acquisition procedure can be found in
Levine et al. (1996). Observed intensities are determined by fitting the photon 
detections to the given positions of the sources listed as ``active'' in the
ASM Source Catalogue. Sources with low ($\leq 2 \sigma$) detections are 
eliminated from the fit and the process is iterated. When an appropriate 
fit is reached, the residuals are searched for new sources, not included 
in the original list. The data from each SSC are analyzed independently. 
Further, data from the three different energy bands ($1.3\, - \,3.0$, $3.0\,
 - \, 5.0$ and $5.0\, - \, 12.2$ keV) are also analysed separately. The 
analysis is performed by the ASM team at the Massachusetts Institute of 
Technology and made publicly available.

The lightcurves for 4U\,0115+634 in the three energy bands, created 
with the software package {\sc xronos}, are presented in Fig. \ref{fig:xray}. 
All the data points used have been obtained from fitted solutions with a 
reduced 
$\chi^{2} \leq 1.5$. Individual dwell data have been distributed into 64800-s 
(18-h) bins. The ASM began operations in January 1996, immediately after the 
end 
of the 1995 Type II outburst.

\begin{figure}[ht]
\begin{center}
    \leavevmode
\epsfig{file=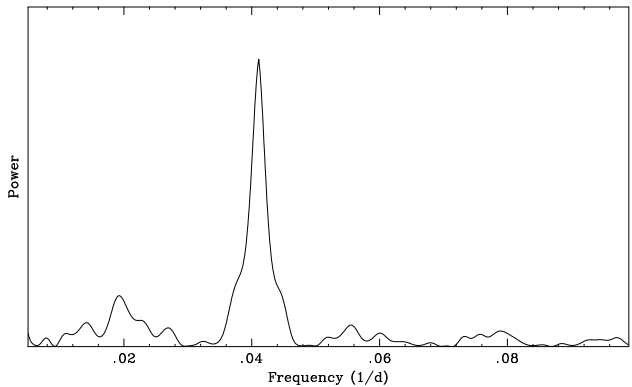, width=8.4cm, bbllx=50pt,
 bblly=190pt, bburx=360pt, bbury=385pt, clip=}
 \end{center}
\caption{Power spectrum for the ASM/{\em RXTE} high-energy band data in the 
interval TJD 10295\,--\,10451, calculated using the CLEAN algorithm (see 
text).}
\label{fig:power}
\end{figure}

\subsubsection{Optical spectroscopy}
\label{subs:vos}

 We managed to observe the source immediately after the 1995 Type II outburst 
and again after the spring gap during which 
the source cannot be observed from La Palma (February -- May). H$\alpha$ 
spectroscopy was taken with the 
Intermediate Dispersion Spectrograph (IDS) on the 2.5-m INT. The telescope was 
equipped with the 235-mm camera + R1200Y
grating which gives a nominal dispersion of 0.8 \AA/pixel. The data
have been processed using the Starlink package {\sc figaro}
(Shortridge \& Meyerdicks 1996) and analysed with the Starlink
package {\sc dipso} (Howarth et al. 1996). Dates of observation and the 
measured equivalent widths (EWs) of the H$\alpha$ line are listed in Table 
\ref{tab:halpha}.

The equivalent width of the spectra are measured by selecting a continuum point
on each side of the line and integrating the flux relative to the straight 
line between the two points using the procedures available in {\sc dipso}. The 
measurements were repeated several times and the error estimated from the 
distribution of values obtained. This error arises due to the subjective 
selection of the continuum. Gaussian fitting to the shapes does not provide a 
better estimate since again the extended wings of the emission line and the 
presence of several atmospheric absorption features make the determination of 
the continuum very imprecise. The errors thus obtained are always $\la 15\%$ 
and typically $\sim 10 \%$. Due to the subjectivity, we prefer to use 15\% as 
a conservative estimate of the error.

\begin{table}[ht]
\caption{Details of the H$\alpha$ spectroscopy for V635 Cas.}
\begin{center}
\begin{minipage}[h]{8.6cm}
\begin{tabular}{lccc}
\hline 
Date of       & TJD & EW of & Line\\
Observation(s)& & H$\alpha$(\AA)\footnote{Errors in EW are 
 $\la 15\%$, due to the subjective continuum determination. See Sect. 
\ref{subs:vos}} & shape\\
\hline
&&\\
Jul. 03, 1995 & 9901 & $-$3.6 & Shell\\
Sep. 12, 1995 & 9972 & $-$4.5 & Shell\\
Jan. 12, 1996 & 10094 & $-$11.3 & Asymmetric\\
Jan. 31, 1996 & 10113 & $-$8.0 & Asymmetric\\
Jun. 20, 1996 & 10254 & $-$7.0 & Asymmetric\\
Jul. 09, 1996 & 10273 & $-$6.5 & Asymmetric\\
Aug. 26, 1996 & 10321 & $-$8.0 & Asymmetric\\
Feb. 01, 1997 & 10480 & $-$3.6 & Asymmetric\\
\end{tabular}
\end{minipage}
\end{center}
\label{tab:halpha}
\end{table}

\begin{figure}[ht]
\begin{center}
    \leavevmode
\epsfig{file=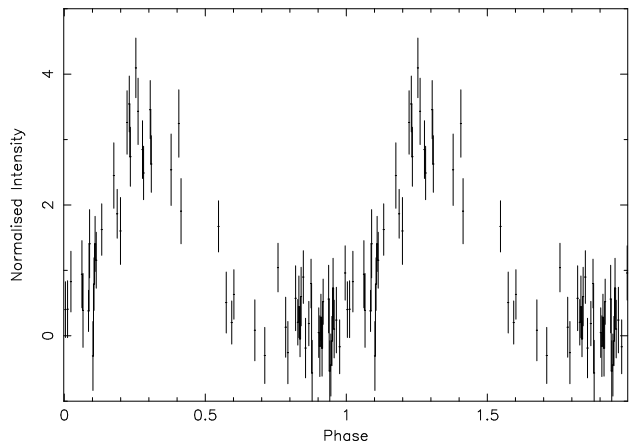, width=8.4cm, bbllx=85pt,
 bblly=165pt, bburx=425pt, bbury=405pt, clip=}
 \end{center}
\caption{Epoch folding of the high-energy ($5.0\, - \, 12.2$ keV) data from 
the All Sky Monitor on board {\em RXTE} for the Be/X-ray transient 
4U\,0115+634 at the orbital period of the neutron star. Phase $\phi = 
0$ is defined as the epoch of periastron passage on TJD 10300.36.}
\label{fig:efold}
\end{figure}

\subsubsection{Infrared Photometry}

Infrared photometry of V635 Cas was obtained with the Continuously Variable 
Filter (CVF) on the 1.5-m Carlos Sanchez Telescope (TCS) at the Teide
Observatory, Tenerife, Spain. Data for the period under discussion are listed 
in Table \ref{tab:irmag}. 

\begin{table*}[ht]
\caption{Observational details of the IR photometry for V635 Cas.}
\begin{center}
\begin{tabular}{lcccc}
\hline 
Date of       & TJD & J mag & H mag & K mag\\
Observation(s)& &  & &\\
\hline
&&\\
Aug 02, 1995 & 9931 &12.15$\pm$0.08 &11.45$\pm$0.05  &11.08$\pm$0.07\\
Oct 15, 1995 & 10005 & 10.81$\pm$0.04 & 10.21$\pm$0.04 &  9.78$\pm$0.040\\
Jan 12, 1996 & 10094 & 11.36$\pm$0.05 & 10.74$\pm$0.05 &10.35$\pm$0.05\\
Jul 28, 1996 & 10293 & 11.50$\pm$0.04 & 10.89$\pm$0.04 & 10.65$\pm$0.04
\end{tabular}
\end{center}
\label{tab:irmag}
\end{table*}

Our data show that, as in previous occasions, the Type II X-ray outburst was 
preceded by the sudden brightening of the infrared magnitudes of V635 
Cas, with an increase of $\approx$ 1.3 magnitudes in $K$ in $\sim$ two months
(August\,--\,October 1995), but the source faded quickly after the outburst.

\subsection{Observations of A\,0535+262 (HD 245770)}

We re-analysed the spectroscopic 
dataset for HD 245770 presented by 
Cl98. All the raw observations from the period 1990\,--\,1993 
were extracted from the original tapes when available or retrieved from the La 
Palma Archive (Zuiderwijk et al. 1994) and reprocessed. This allowed us to 
find 
some new spectra which had not been included in the work of Cl98. The dates of 
observation for the complete set of spectra are listed in Table 
\ref{tab:nosimon}, together with measurements of the equivalent width (EW) of 
H$\alpha$. The spectra are displayed in Fig. \ref{fig:global}. The method
followed to measure the EW was the same as for V635 Cas, and the comments 
on the uncertainty of the measurement also apply.

\begin{table}[ht]
\caption{Details of the H$\alpha$ spectroscopy for HD 245770.}
\begin{center}
\begin{minipage}[h]{8.6cm}
\begin{tabular}{lccc}
\hline 
Date of       & TJD &Telescope & EW of \\
Observation(s)&     &      & H$\alpha$(\AA)\footnote{Errors in EW are 
 $\la 15\%$, due to the subjective continuum determination. See Sect. 
\ref{subs:vos}}\\
\hline
&&&\\
Feb 7, 1990 & 7929 & INT & $-$12.4\\
Feb 21, 1990 & 7943 & INT & $-$10.9\\
Apr 9, 1990 & 7990 & INT & $-$10.9\\
Nov 14, 1990 & 8209 & INT & $-$9.9\\
Dec 27, 1990 & 8252 & INT & $-$9.0\\
Jan 28, 1991 & 8284 & INT & $-$8.8\\
Apr 16, 1991 & 8362 & WHT & $-$7.1\\
Aug 28, 1991 & 8496 & INT & $-$8.0\\
Dec 13, 1991 & 8603 & INT & $-$10.6\\
Feb 18, 1992 & 8670 & PAL & $-$10.6\\
Aug 17, 1992 & 8851 & PAL & $-$7.5\\
Aug 18, 1992 & 8852 & PAL & $-$7.3\\
Mar 8, 1993 & 9054 & PAL & $-$13.6\\
Mar 10, 1993 & 9056 & PAL & $-$13.6\\
Sep 23, 1993& 9253 & PAL & $-$10.3\\
Dec 5, 1993 & 9326 & PAL & $-$14.0\\
Dec 6, 1993 & 9327 & PAL & $-$13.7 \\
\end{tabular}
\end{minipage}
\end{center}
\label{tab:nosimon}
\end{table}
 
The spectra displayed in Fig. \ref{fig:growth} are taken from Cl98, except 
for the 1994 March 25 spectrum, which was obtained with the the JKT equipped 
with the R1200Y grating and has lower dispersion ($\approx$ 1.2 \AA/pixel).
These spectra were taken with the 2.6-m telescope at the Crimean Astronomical
Observatory and have not been re-reduced (see Table 1 in Cl98 for details).

\subsection{Observations of V\,0332+53 (BQ Cam)}

Observations of BQ Cam, the optical counterpart to the Be/X-ray transient 
V\,0332+53 have been carried out during our campaign. In this paper, we only 
report on H$\alpha$ observations obtained during the period 1990\,--\,1991. 
These spectra were obtained with the IDS on the INT and different gratings, 
generally the 
R1200Y, and are displayed in Fig. \ref{fig:vrbq}. Further discussion of the 
properties of this system is left for a forthcoming paper (Negueruela et al., 
in preparation).

\section{Results}
\subsection{4U\,0115+634}

The X-ray lightcurve of 4U\,0115+634 (Fig. \ref{fig:xray}) between January 
and July 1996 (TJD 10087\,--\,10294) is compatible 
with the complete absence of emission (no detections above the 2$\sigma$ 
threshold according to the quick-look results provided by the {\em RXTE}/ASM 
team). However, starting in early August 1996, a succession of outbursts is 
clearly visible. 
The first outburst was also seen by BATSE (Scott et al. 1996), which on August 
12, 1996 (TJD 10307) observed a pulsed flux in the 20\,--\,50 keV range of 
$\sim 30$ mCrab. Two other outbursts were weakly detected by BATSE (Finger, 
1997, priv. comm.). These two outbursts are clearly visible in the 
ASM/{\em RXTE} lightcurve. There is indication of  
a weak outburst around September 4, 1996 (TJD 10330, one orbital period after 
the first outburst).

\begin{figure*}[ht]
\begin{center}
    \leavevmode
\epsfig{file=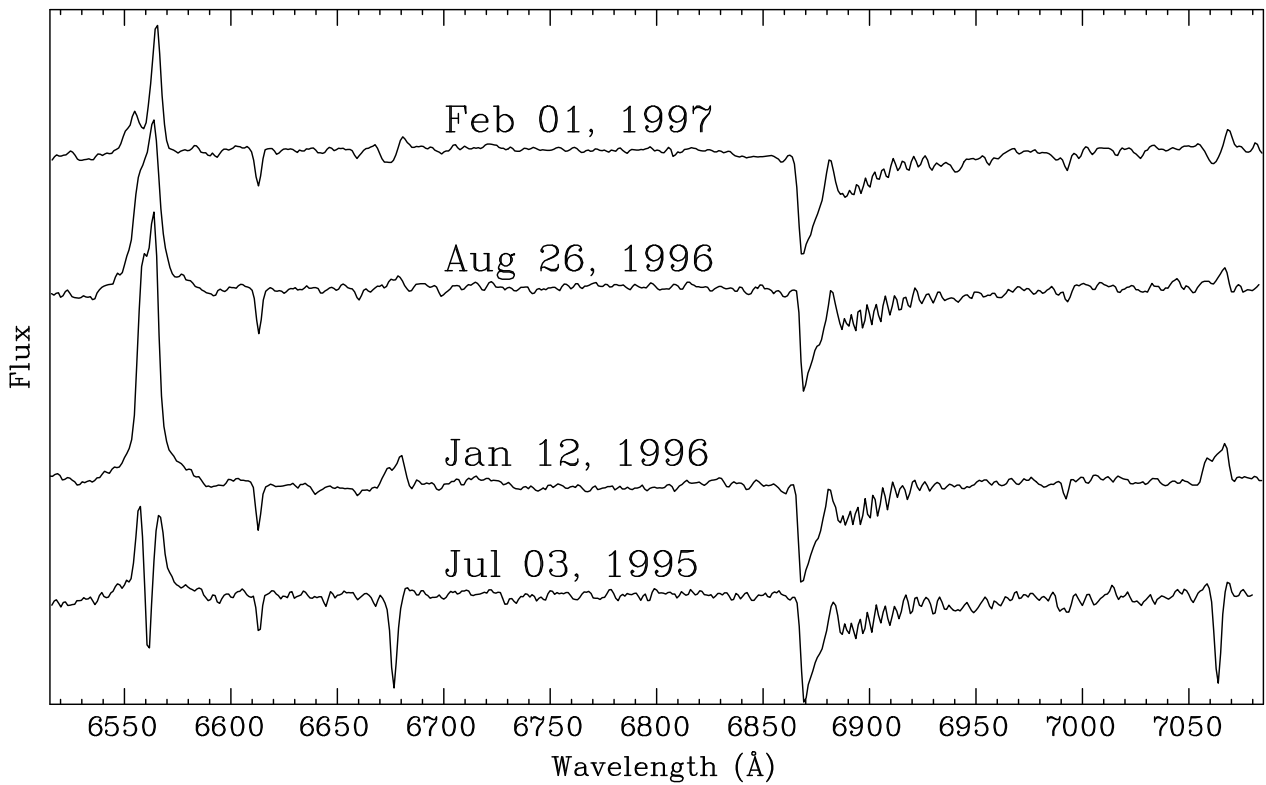, width=16.8cm, bbllx=50pt,
 bblly=330pt, bburx=450pt, bbury=585pt, clip=}
 \end{center}
\caption{Series of spectra showing the appearance of a global density 
perturbation, eventually leading to a global density wave, in the envelope
of the Be/X-ray binary 4U\,0115+634. Notice the parallel evolution of 
the He\,{\sc i} $\lambda \lambda$6678, 7065 \AA\ lines. The spectra have been 
divided by a spline fit to the continuum for normalisation and arbitrarily 
offset for clarity. The July 1995 spectrum is
typical of the quiescent state of the source.}
\label{fig:developo}
\end{figure*}

The outbursts are clearly seen in the high-energy ($5.0\, - \, 12.2$ keV) band,
but hardly detectable in the low-energy band. Due to this large hardness ratio,
in the following analysis we have only used the high-energy band data. The 
periodogram of the X-ray lightcurve was calculated using the CLEAN procedure 
from the Starlink {\sc period} package (Dhillon \& Privett 1997), with 10 
iterations and a gain of 0.1 in each step (see Fig. \ref{fig:power}. 
There is a single dominant peak corresponding to a period of 
24.4$\pm$0.1 d, the orbital period of the neutron star. This result is also 
obtained when applying other period searching procedures such as SCARGLE 
(Lomb-Scargle normalised periodogram) or FT (discrete Fourier power spectrum).
This is the first occasion in which the X-ray emission of 4U\,0115+634
has displayed modulation of any kind. All the outbursts previously observed 
since the discovery of the source in 1969 had been of Type II and had not 
displayed any orbital modulation.
The periodicity is not observed when only data from before TJD 10295 or after 
TJD 10450 are analysed, confirming that the modulation is due to 
the outbursts and not to any quiescence emission. A Fisher 
randomization test shows that no peak has a probability $\geq$ 60\% of being
real and there are no peaks in the range 20\,--\,30 days.

The high-energy band ASM data for the time TJD 10295\,--\,10451 (the period of 
X-ray activity) were folded at the orbital period, using 
$P_{{\mathrm orb}} = 24.32 \: {\mathrm d}$ and epoch of periastron passage TJD 
10300.36 after the model in N97. The folded lightcurve is shown in Fig. 
\ref{fig:efold}. The outbursts are seen to peak close to orbital phase $\phi 
\sim 0.3$, far away from periastron. 

The optical counterpart, which had reached a peak in brightness just before
the outbursts, fainted steadily  during this period, as can be seen in Table
\ref{tab:irmag}. 
As indicated in N97, it is difficult to define a photometric `quiescent' 
state for this source. The $J$ magnitude oscillates between $\sim 
10.8$ and $\sim 12.3$. Whenever it has come close to $J \la 11$, an X-ray 
outburst has taken place. The infrared colours remain relatively constant, as 
is usual in the system (N97), except for the July 1996 observations.

\subsection{A\,0535+262}
\label{sec:rhd}

\begin{figure}[ht]
\begin{center}
    \leavevmode
\epsfig{file=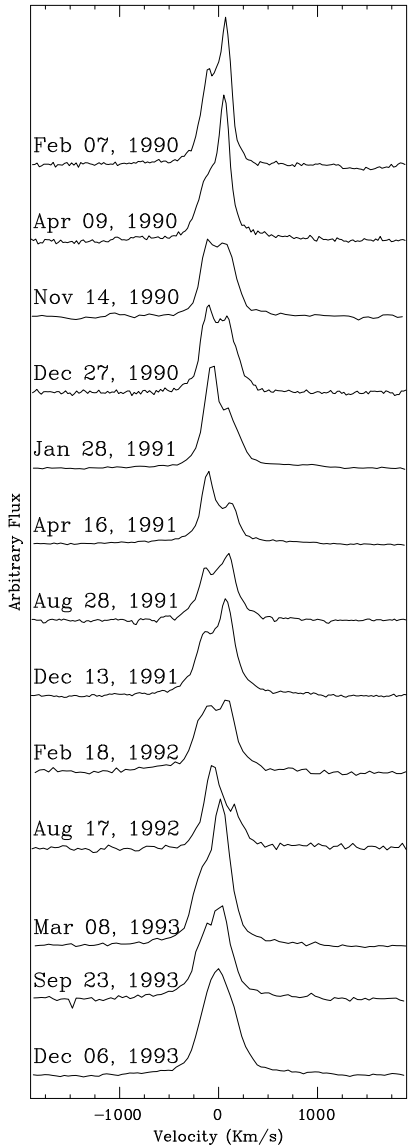, width=5.8cm, bbllx=95pt,
 bblly=35pt, bburx=285pt, bbury=535pt, clip=}
 \end{center}
\caption{Evolution of the shape of the H$\alpha$ line in HD 245770 
during 1990\,--\,1993. The observed profiles indicate the presence of a global
oscillation with a quasi-period of $\sim$ 18 months. All the spectra have been
smoothed with 
a Gaussian function ($\sigma = 0.8$ \AA) to obtain a comparable resolution, 
divided by a spline fit to the continuum for normalisation and offset for 
display.}
\label{fig:global}
\end{figure}

The spectra displayed in Fig. \ref{fig:global} show that HD 245770 was 
displaying quasi-cyclic V/R variability during the period 1990 -- 1993. As 
noted by Cl98, the shape of the spectra is too complex to attempt a Gaussian 
fitting to the profiles. In many spectra, very extended wings are apparent, 
which makes the determination of the continuum 
very imprecise. However, an attempt has been made to use the same 
criteria for all the spectra, so that the values measured on different spectra 
can be compared. Our values are in general agreement with those of Cl98. The 
intrinsic inaccuracy of the measurements, together with the reduced number of 
spectra, precludes the possibility of a proper search for periodicity. In 
spite of 
this and of the incompleteness of the coverage, due principally to the gap 
during which the source cannot be observed from the ground (May -- July), a 
quasi-period of $\sim 18\pm 1$ months can be deduced from visual inspection. 
It is noteworthy that this period is close to the 508-d period 
detected by Hao et al. (\cite{hao}) in the photometric lightcurve of the 
source during the same epoch. 
Clark et al. (\cite{clarkb}) argue that this modulation does not seem to be 
coherent 
over long time-scales, but this behaviour is what should be expected of the 
quasi-periodicity of V/R cycles.

\begin{figure}[ht]
\begin{center}
    \leavevmode
\epsfig{file=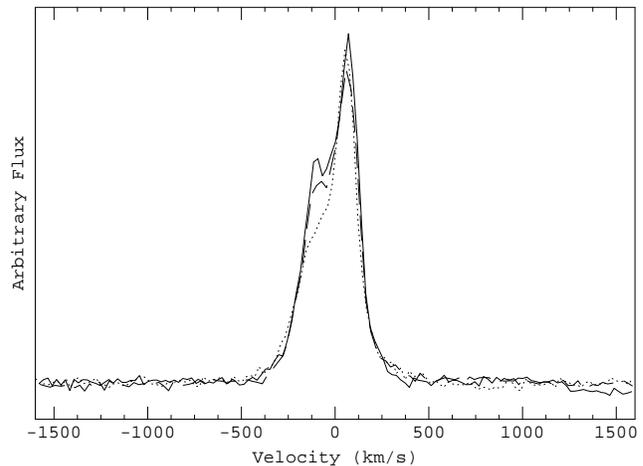, width=8.4cm, bbllx=65pt,
 bblly=215pt, bburx=420pt, bbury=475pt, clip=}
 \end{center}
\caption{Evolution of the H$\alpha$ line profile in HD 245770 during February 
-- April 1990. The slow decrease in the strength of the blue arm is typical of 
the speed of variation seen during the four years before the Type II outburst. 
The 
solid line represents the spectrum from February 7, the dash line, that from 
February 21 and the dotted line the spectrum from April 9. All 
the spectra have been normalised and offset by a constant amount.}
\label{fig:slowo}
\end{figure}

\begin{figure}[ht]
\begin{center}
    \leavevmode
\epsfig{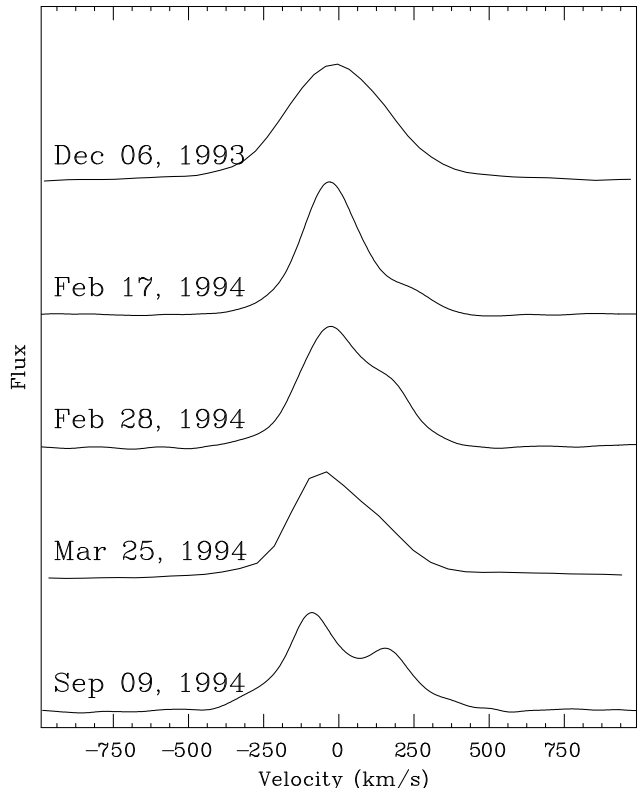}
 \end{center}
\caption{Evolution of the H$\alpha$ line profile of HD 245770 around 
the time 
of the Type II X-ray outburst peaking on February 18, 1994. The growth of a 
red shoulder in about ten days is clearly seen. All the spectra have been 
normalised and offset by a constant amount. The March 25 spectrum was taken 
with 
the JKT and has lower resolution. The higher resolution spectra have been 
smoothed with a $\sigma$ = 0.8 \AA\ Gaussian function.}
\label{fig:growth}
\end{figure}

The continuation of the cyclic behaviour would imply a blue-dominated profile 
during early 1994, similar to those observed in January 1991 or August 1992. 
As can be seen in Fig. \ref{fig:growth}, the cyclic behaviour was broken in 
coincidence with the Type II outburst. Between February 17 and February 28 a 
strong red shoulder formed, changing the global shape of the emission line. 
This 
change was much faster than the variations associated with the cycle, as 
can be seen comparing Fig. \ref{fig:growth} with Fig. \ref{fig:slowo}, and 
must be of a different nature.

\subsection{V\,0332+53}
\label{sec:rbq}

\begin{figure}[ht]
\begin{center}
    \leavevmode
\epsfig{file=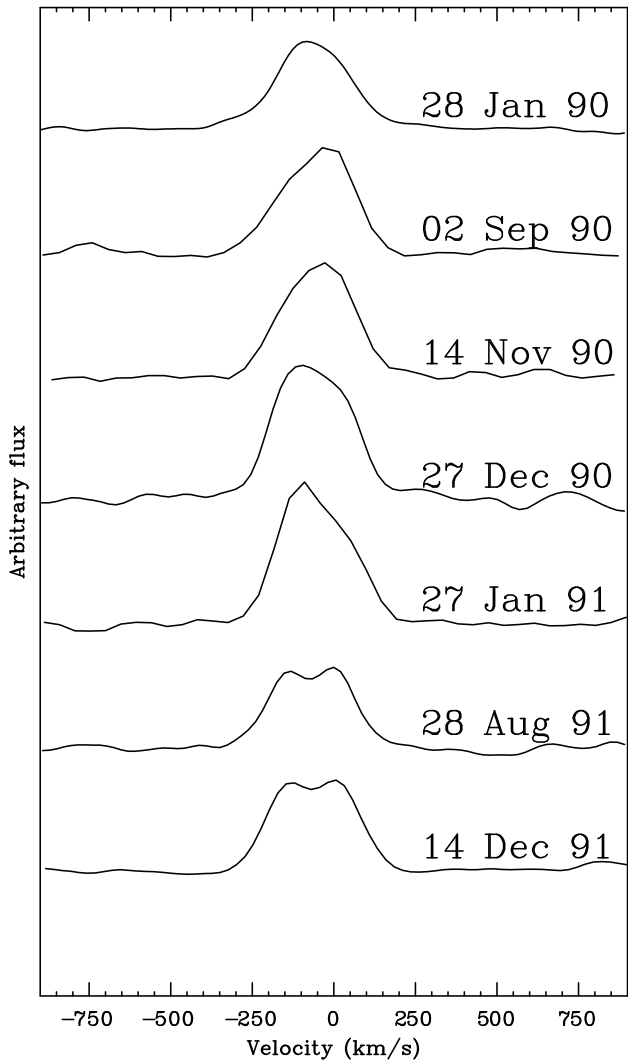, width=8.4cm, bbllx=130pt,
 bblly=135pt, bburx=420pt, bbury=620pt, clip=}
 \end{center}
\caption{H$\alpha$ observations of BQ Cam, the optical counterpart
of V\,0332+53 during the period 1990 --- 1991, showing the propagation
and final disappearance of a density perturbation. The spectra have been 
divided 
by a spline fit to the continuum for normalisation.}
\label{fig:vrbq}
\end{figure}

Figure \ref{fig:vrbq} shows that V/R 
variability was present in BQ Cam during 1990, with a quasi-period of $\sim$ 1 
year, but it had disappeared by 1992. The variations in the lines are smaller 
than in the case of HD 245770, presumably because the star is seen almost 
pole-on. The mass function for the system implies $i < 15^{\circ}$ (Corbet et 
al. 1986a) and the orbit is expected to be co-planar with the equatorial disc 
(Waters 
et al. 1989). 
No X-ray emission has been detected from the source since late 1989 (Bildsten 
et 
al. 1997). 
The strength of H$\alpha$ emission seems
to have remained approximately constant, indicating that
the cessation of the V/R variability was not associated with any major
change in the size of the disc.

\section{Discussion}
\label{sec:discu}

\subsection{A global change in 4U\,0115+634}

For the first time since its discovery in 1969, 4U\,0115+634 has been
observed to undergo a series of X-ray outbursts modulated with the orbital 
period. The outbursts did not peak near periastron but at phase $\sim 0.3$.
The fact that we obtain such a strong modulation of the X-ray lightcurve at the 
orbital period with only three outbursts argues against the idea that the 
delay in the peak of the outbursts can be due to the mediation of an extended 
accretion disc, which would not be so regular. 
A more likely explanation would be that the neutron star 
repeatedly went through a region of 
very high density at some point of its orbit close to phase 0.3. The spectra 
from the summer of 
1996 clearly indicate that the series of Type I outbursts was not associated 
with any major change in the size of the disc (as reflected in the EW). The 
infrared magnitudes of the source were close to its quiescence magnitudes 
during the first (and strongest) Type I outburst. All previous outbursts had
been associated with much brighter infrared magnitudes (at least half a 
magnitude brighter than the August 1996 values). 
The only observed difference with previous quiescence states is the presence 
of asymmetric emission line profiles, which are indicating the existence of an 
asymmetric density configuration. 

The spectrum of the source underwent a major change during or immediately
before the December 1995 Type II outburst. The usual quiescence shell spectrum 
(see N97), 
which was observed only two months before the outburst, was replaced by strong 
asymmetric emission lines  (see Fig. \ref{fig:developo}). The asymmetric 
profiles, characterised by a dominant red peak and a smaller blue peak were 
still present more than a year later, indicating that the circumstellar 
envelope has been strongly disturbed.

\begin{figure}[ht]
\begin{center}
    \leavevmode
\epsfig{file=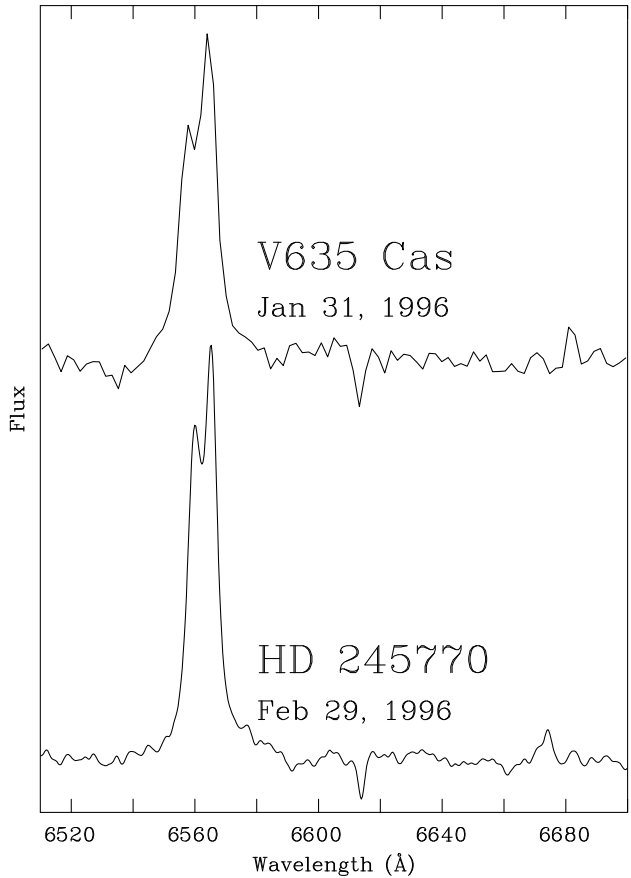, width=8.4cm, bbllx=115pt,
 bblly=175pt, bburx=405pt, bbury=575pt, clip=}
 \end{center}
\caption{Comparison of the H$\alpha$ line profile in V635 Cas and
in  HD 245770 (A\,0535+26) after a giant X-ray outburst. The HD 245770 
spectrum was taken with the 1.0-m JKT in La Palma on Feb. 28, 1996, while 
the star was undergoing a V/R cycle. The
spectra have been smoothed with a Gaussian filter of $\sigma = 0.8$
\AA\  and had their continua normalised. Note that both stars are
seen under a similar moderate ($\approx 40^{\circ} - 60^{\circ}$)
inclination angle.}
\label{fig:lateev}
\end{figure}

Fast changes in the line profiles are not rare in this source (N97). 
However, this is the first time in which strongly asymmetric profiles, 
indicating a perturbed density distribution, have been observed. 
The asymmetric spectra of V635 Cas are very similar to those of V/R 
variable systems such as HD245770 or LS\,I$\:$+61$^{\circ}$235 
(Reig et al. 1997a). Figure \ref{fig:lateev} shows the similarity of the 
H$\alpha$ line profiles of V635 Cas and HD 245770 when their 
circumstellar envelopes have been perturbed. 

The previous absence of Type I outbursts is explained by  
centrifugal inhibition of accretion  due to the fast rotation and strong 
magnetic field of the neutron star (N97). The centrifugal barrier can only 
be overcome if the ram pressure of incoming material rises (SWR). The increase 
in ram pressure can be due to a higher density of the surrounding material or a 
higher relative velocity between the neutron star and the environment. The 
spectra do not show any evidence for enhanced mass loss during August 1996, but
indicate that different regions in the envelope have different densities. This
hints strongly at the possibility that the Type I outbursts are associated with
the presence of a region of enhanced density in the envelope.

\subsection{Fast V/R variations in A\,0535+262}

Between 1990\,--\,1993, HD 245770 was showing V/R variability with a 
quasi-period of $\sim 18$ months. 
Even though small changes in the EW of H$\alpha$ are expected 
on short time-scales (since they are observed for most Be stars), there seems 
to 
be a cyclic variation associated with the V/R cycle, in the sense that the 
red-peak phases (or perhaps the symmetric phase that immediately follows them) 
show larger EWs than other phases of the cycle. This seems to be superimposed 
on 
a general trend: the values of the EW during the second observed cycle are 
systematically smaller than in the first cycle, but those from 1993 (third 
cycle) are consistently higher than in the previous years, which in the 
standard 
interpretation would indicate 
an increase in the size of the disc. It is noticeable that the source started 
its Type I X-ray activity during 1993. Unfortunately, there is a gap in the 
observations in the second half of 1992 that does not allow us to determine if 
the increase in the EW of H$\alpha$ was gradual or a fast event. In any case, 
the V/R cycle does not seem to have been strongly affected by this (compare the 
very similar shapes of the line profiles in April 1990 and March 1993, which 
are 
separated by two complete cycles). Therefore the conditions in the disc must 
have remained relatively constant until the Type II outburst in February 1994.

As indicated in Sect. \ref{sec:rhd}, the change in the shape of H$\alpha$ 
that 
took place during the outburst was much faster than any changes associated with 
the cyclic V/R variability. It could not have been due to the existence of an 
accretion disc around the neutron 
star, because the structure of the line was basically unchanged a month later, 
when the X-ray outburst had finished and the neutron star was in a very 
different orbital position. 
Therefore the growth of the red shoulder must have reflected a global change in 
the structure of the disc, which took place as the same time as the Type II 
X-ray outburst.

This is confirmed by the fact that the spectra from late 1994 correspond to 
approximately the same phase of the V/R cycle as those at the beginning of the 
year (compare the spectrum from February 28 with that from September 9). Cl98 
indicate that the quasi-period of V/R variability after September 1994 was 
approximately one year. The process that took place at the same time 
as the Type II outburst resulted in a change of both the period and the phase 
of 
the quasi-cycle. Since these parameters depend strongly on the density gradient 
in the disc, it must have implied a major perturbation of the 
physical conditions in the disc.

\subsection{V/R cycles in other Be/X-ray binaries}
\label{sec:vrvar}

Some Be/X-ray binaries have been known to display V/R variability for many 
years, showing quasi-periods in the same range as those observed in isolated 
Be stars. The Be star $\gamma$ Cas, extensively studied over the years, is 
the optical component of the X-ray source 2S\,0053+604, which is 
believed to contain an accreting white
dwarf (Haberl 1995). It has shown V/R variability in many occasions, with 
quasi-periods between 4 and 7 years (Doazan et al. 1987). Since 1970, the V/R  
cycle has had a period of 5$\pm$1 years (Telting et al. 1993). Likewise, X 
Persei, the optical counterpart to 4U\,0352+309, has been observed to 
undergo phases of 
V/R variability with quasi-periods ranging from 2 to 12 years (see Okazaki 
1997 for references). These sources are not transients, but persistent 
low-luminosity X-ray emitters and it is believed that the orbits of the 
compact objects are very wide. Another source displaying V/R variability is 
LS\,I$\:$+61$^{\circ}$\,235, 
the optical component of the Be/X-ray binary RX\,J0146.9+6121. It 
shows a quasi-period of about three years (Reig et al. 1997a). This source
was only discovered in 1991 and it is not certain yet whether it is a 
transient or a persistent source, though the second possibility seems 
more likely. 
Given the very wide orbits of the neutron star in these objects, we have no 
reason to suspect that their V/R variability is different at all from that 
seen in isolated Be stars.

Among the transients, BQ Cam, the optical component of 
V\,0332+53, displayed V/R 
variability on a time-scale of weeks or a few months during a series of Type I 
outbursts in 1983 (see Corbet et al. 1986a and references therein). As shown 
in Sect. \ref{sec:rbq}, it was displaying quasi-cyclic V/R variability soon 
after 
the 1989 Type II outburst. V801 Cen, the optical component of the 
southern Be/X-ray binary 4U\,1145$-$619 showed large variability in 
both H$\alpha$ and H$\beta$ in one week during an X-ray outburst in January 
1985 
(Cook \& Warwick 1987). Long-term V/R variability has also been observed in 
this 
object. Stevens et al. (\cite{stevens}) present data that could be explained 
by the existence of a quasi-cycle with a period $\sim 3$ years. Further 
observations confirm both the existence of quasi-cyclic V/R variability and 
faster variations (Stevens, 1997, priv. comm.).

There is no reason to believe that the 
behaviour of this perturbation cannot be explained by the theory of one-armed 
global oscillations. As in the case of 4U\,0115+634 and 
A\,0535+26, the profile shapes observed during these slow quasi-cyclic 
variations
are not distinguishable from those seen during periods of fast variability.

\subsection{Interpretation}

We have presented observational evidence that most Be/X-ray binaries 
(persistent and transient sources) display V/R variability with 
quasi-periods which are not 
correlated at all with their orbital periods. These observations invalidate 
the model of Apparao \& Tarafdar (1986), who suggested that the V/R 
variations in Be 
binaries were due to the presence of an emission region associated with the 
Str\"{o}mgren sphere of the neutron star and should therefore be modulated 
with the orbital period.

The only case in which some evidence could support this model is 
4U\,1258$-$61 (GX\,304$-$1). Corbet et al. (\cite{corbetb}) 
obtained spectroscopy of its optical counterpart, V850 Cen,
between 1977 and 1983, during its active Be (and X-ray) phase. They observed 
V/R variability with a quasi-period of
approximately 130 days, which is very close to the orbital period
of the neutron star in the system (132.5$\pm$0.5 d). Their statistical 
analysis 
showed that the possibility that the V/R ratio was modulated at the orbital 
period is $\ga$ 85\%. However, our monitoring 
reveals no evidence at all of any modulation in the shape or strength of the 
emission lines with the orbital period in any of the Be/X-ray binaries 
included in the programme. The observed V/R 
variability can be readily explained with the existing models of one-armed 
oscillations developed for isolated Be stars. We believe that, if the 
coincidence between the orbital period and the V/R quasi-period reported for 
4U\,1258$-$61 is real, it is 
more likely to be explained by some kind of resonance than by the Apparao \& 
Tarafdar model.

We have presented observational evidence that large density perturbations
arose in the envelopes of the Be/X-ray binaries 4U\,0115+634 and 
A\,05353+26 in coincidence with Type II outbursts and that they 
strongly affected the dynamics of their envelopes. In the case of 
4U\,0115+634, the change from 
symmetric emission lines to asymmetric profiles has resulted in the 
commencement 
of V/R variability. In the case of A\,0535+26, the existing pattern of 
variability was profoundly affected, with a shift in both the quasi-period and 
phase of the V/R cycle. 

In both cases, the fast appearance of the density disruption 
during the X-ray outburst suggests that there is an association between the 
Type II outburst and the density perturbation. No isolated Be stars have ever 
been 
observed to go from symmetric profiles to clearly asymmetric ones in such a 
short time-scale. Moreover, this kind of fast variations has been seen to 
occur only during X-ray outbursts, when the neutron star is closer to the 
envelope, which again points to a connection between both events.

However, it must be noticed that we do not know how rapidly an instability 
develops into a global mode in an isolated Be star, but there are indications
that it can be quite fast -- less than a year in the Be star 66 Oph
(Hanuschik et al. 1995). The direction of the causal connection between the 
two events, if any, cannot be deduced from the observations. It could well be 
that 
both the density perturbation{\em and} the outburst were caused by a common 
cause, such as violent asymmetric ejection of material from the Be star.

On the other hand, the association between Type II X-ray outbursts and major 
changes in the circumstellar envelopes of Be/X-ray binaries seems to be clear. 
Stevens et al. (\cite{stevens}) argue that a giant outburst of 
4U\,1145$-$619 in March 1994 was followed by a reduction in the 
strength of H$\alpha$ by a factor of two. A similar reduction of the strength 
of 
H$\alpha$ (by at least 30 \%) took place after an X-ray outburst of 
A\,1118$-$616 (Coe et al. 1994). The new observations confirm that the 
changes in the circumstellar envelopes at the time of Type II X-ray outbursts 
are global and profound, affecting their basic physical properties. The fact 
that the red shoulder in the H$\alpha$
profile of A\,0535+26 seems to have grown {\em after} the start of the
X-ray outburst (the X-ray outburst peaked on February 18, while the spectrum 
taken just two days before shows the red shoulder beginning to appear) 
suggests that it is the X-ray outburst -- or some physical process associated
with it -- which affects the dynamics of the envelope rather than the 
opposite.

Several objects have now been observed to display fast V/R variations (with a 
time-scale of days) and cyclic V/R variability with a time-scale of months. In 
a one-year V/R cycle, a phase such as the disappearance of the red peak 
illustrated in Fig. \ref{fig:slowo}, would take about one month, if we can 
assume that the whole process happens at the same speed. The formation of a 
red shoulder seen in Fig. \ref{fig:growth} was only a factor 3\,--\,4 faster 
than it would be in a normal V/R cycle. There is not an evident separation 
between the time-scales of fast changes
and slow quasi-cyclic variability. Moreover, the shapes observed during these 
fast variations are not distinguishable from those observed during the V/R 
cycles. Therefore there is no longer any reason to suspect that the fast 
variations in the emission line 
profiles during Type II outbursts are caused by any other physical processes 
different from those involved 
in quasi-cyclic V/R variability. We propose that both types of variability are 
associated with highly non-axisymmetric density distributions in the envelopes. 
The rapid changes in the line profiles can be due to strong disruptions of
the distribution of material in the envelopes taking place during very short 
time-scales, while the quasi-cyclic variability is due to the propagation
of density waves in the discs, which originate slowly-moving disruptions. It is 
even possible that the 
global disruption events can act as original excitations giving rise to 
the global modes.

The observations of Hanuschik et al. (\cite{han95}) and Hummel \& Vrancken 
(\cite{hummel95}) support the idea that the density perturbations 
in isolated Be stars expand outwards through the discs, from the neighbourhood
of the central star. These perturbations would cover the whole radial span 
over a typical time-scale of the order of one year, in agreement with the
calculations of Okazaki (\cite{oka91}). In the Type II outbursts of Be/X-ray 
binaries, the appearance of asymmetry in the H$\alpha$ emitting region
and the X-ray outburst -- indicating a perturbation at the distance of the 
neutron star orbit -- are almost simultaneous.

It must be stressed that all known Be/X-ray transients which have displayed V/R
variability show very short quasi-periods of V/R variability, compared to 
isolated Be stars and to the predictions of one-armed oscillation models. 
Be/X-ray transients 
for which complete cycles have been observed are V850 Cen (quasi-period 
$\approx 4$ months), HD 245770 (about one year), BQ Cam (about 
one year) and V801 Cen
(about three years). In isolated Be stars, periods range from 2 years to 
decades, with an average of 7 years. Okazaki (\cite{oka91}) found that, in 
isolated Be stars, the periods of the oscillations are larger for larger discs 
or smaller density gradients in the discs. Waters et al. (\cite{waters88}) 
found 
that the density gradients in the envelopes of Be/X-ray binaries were in the 
same range as those calculated for isolated Be stars. 

Recently, Okazaki (\cite{oka97}) has suggested that the higher radiation 
pressure could induce shorter quasi-periods for the earliest Be stars, but the 
observed quasi-periods in Be/X-ray binaries are still too short. Okazaki 
(1997, priv. comm.) suggests that the shorter periods could be due to 
denser envelopes. This 
systematic difference supports the idea that the presence of an orbiting 
neutron 
star is a major factor affecting the dynamics of the extended circumstellar 
envelopes of 
Be/X-ray transients, as has been proposed by Reig 
et al. (\cite{reigb}) in order to explain the correlation between the {\em 
maximum}\  EW of H$\alpha$ ever observed from a Be/X-ray binary and its orbital 
period. They suggested that the correlation exists because the continuous 
passage of the neutron star prevents the development of a large circumstellar
disc by accreting material at periastron passage. A second
possibility could simply be that the presence of the neutron star
makes the discs unstable against density perturbations and the presence
of these perturbations prevents their further growth. 

An interesting point is the fact that the only series of Type I outbursts from 
4U\,0115+634 
ever observed seems to be associated with the presence of a density 
perturbation in its envelope. This suggests the possibility that the existence 
of moving regions with enhanced 
densities and perturbed velocity fields causes the
series of Type I outbursts observed from close Be/X-ray transients. The 
Be/X-ray 
transient 2S\,1417$-$624 showed a series of Type I outbursts in 1995, 
after a 
giant outburst in late 1994 (Finger et al. 1996b), mimicking the behaviour of 
4U\,0115+634. The Type I outbursts peaked close to apastron, again 
suggesting 
that the density distribution was not symmetric. This behaviour would be 
easily explained if a global density perturbation in the envelope had been 
started at the time of the Type II outburst and the Type I outbursts were 
caused 
by the passage of the neutron star through the perturbed region. This 
explanation would also account for the large changes in relative velocity
between the neutron star and the material in the envelope needed to explain 
the X-ray lightcurves of different outbursts without having to invoke
enormous variations in the outflow rate from Be stars, which do not
seem to be reflected in the observations. All the observed profiles can be 
explained by Keplerian movements, without any indication of mass outflow.

The main unknown is the extent of the H$\alpha$ emitting region. 
Okazaki (\cite{oka97}) has suggested that the global oscillations are confined 
to the inner parts of the discs due to the effect of radiation pressure. In 
most of his models, the perturbations would not extend to the distance of 
periastron passage of the companion. However, Hummel \& Hanuschik 
(\cite{hummel97}) have shown that 
the existing models, though qualitatively explaining the main characteristics 
of 
V/R variability, cannot be used to obtain accurate estimates of the perturbed 
density and velocity fields. The approximation of linear perturbations used in 
all existing models cannot reproduce the strength of asymmetry in observed 
profiles. If the 
perturbations could reach the distance at which the neutron star approaches 
(typically 8 -- 12 R$_{*}$ for the close-orbit transients), the series of Type 
I 
outbursts could be easily explained. It is evident from the observations that 
most of the H$\alpha$ emitting region is affected by the perturbation. If the 
typical values of the outer radii of H$\alpha$ emitting regions measured by 
Hummel \& Vrancken (\cite{hummel95}) for Be stars (7 -- 30 R$_{*}$) can be 
extrapolated to Be/X-ray binaries, this is a likely possibility.

\section{Conclusions}

We have presented observational evidence showing that global disruptions
are frequent in the extended circumstellar envelopes of Be/X-ray binaries. 
These perturbations are reflected in the asymmetric line profiles normally
observed from these systems. V/R ratio variability is observed to occur with 
typical time-scales ranging from a few days to several years. In at least two
cases (the giant outbursts of A\,0535+26 in February 1994 and 
4U\,0115+634
in December 1995), a major disruption seems to have originated in coincidence
with the X-ray outburst.
Further evidence of the association between fast changes in the line profiles
and X-ray outbursts has been seen in most Be/X-ray transients.

We believe that all these observation suggest that the presence of the neutron 
star represents a major factor controlling the dynamics of the discs around 
the Be stars in X-ray binaries. This fact provides an explanation to the 
correlation between maximum H$\alpha$ EW and orbital period found by Reig et 
al. (\cite{reigb}). The frequent presence of major density perturbations in 
the envelopes of Be/X-ray binaries introduces a new element of complication
in the modelling of these systems. Rather than assuming that the disc is
static and homogeneous, new models should take into account the presence of  
global density waves and explore the possibility that the series of Type I 
outbursts are caused by the interaction of the neutron star with the regions of 
enhanced density which these waves generate.

Continued monitoring of Be/X-ray transients and careful optical 
coverage of future Type II outbursts will provide the only test for these 
hypothesis.

\begin{acknowledgements}

We are grateful to the INT and WHT service programmes for additional 
optical observations. Special thanks to the INT service programme 
(particularly Phil Rudd and Don Pollacco) for the monitoring of V635 Cas.
The JKT, INT and WHT are operated on the island of La Palma by the Royal
Greenwich Observatory in the Spanish Observatorio del Roque de
Los Muchachos of the Instituto de Astrof\'{\i}sica de Canarias. The
1.5-m telescope at Mount Palomar is jointly owned by the California
Institute of Technology and the Carnegie Institute of Washington. We
are very grateful to all astronomers who have taken part in
observations for this campaign,  G. Capilla, D. Chakrabarty, J.S. Clark,
C. Everall, A. J. Norton, A. Reynolds, P. Roche, A.E. Tarasov, J. M. 
Torrej\'{o}n and S. J. Unger. IN would like to thank Atsuo Okazaki for many 
useful comments. The data reduction was mainly carried out using 
the Southampton University Starlink node, which is funded by PPARC. This 
research has made use of the La Palma Data Archive and of the Simbad database,
operated at CDS, Strasbourg, France.

\end{acknowledgements}

\end{document}